\documentclass{Interspeech}

\interspeechcameraready

\title{Enhancing Acoustic-to-Articulatory Inversion with Multi-Target Pretraining for Low-Resource Settings}

\author{Jesuraj}{Bandekar}
\author{Prasanta Kumar}{Ghosh}

\affiliation[nocounter]{Department of Electrical Engineering}{Indian Institute of Science, Bengaluru}{India}

\email{jesurajbandekar.661@gmail.com, prasantag@gmail.com}


\usepackage{comment}
\usepackage{xcolor}
\usepackage{multirow}
\usepackage{graphicx}

\begin{document}

\maketitle

\begin{abstract}
Acoustic-to-Articulatory Inversion (AAI) estimates vocal tract articulator movements from speech, benefiting tasks like ASR, speech synthesis, and speaker verification. While deep learning-based methods (CNNs, RNNs, Transformers) have advanced AAI, recent studies show that Self-Supervised Learning (SSL) features further enhance performance, particularly in low-resource settings. However, SSL feature extractors introduce inference latency and computational overhead. To address this, we propose a novel pretraining method leveraging three target representations—Phoneme Labels, Articulatory Feature Labels, and Critical-articulator Labels—eliminating the need for an SSL extractor during inference. We evaluate our approach against both baseline and SSL-based models across various data conditions. Results demonstrate that our method consistently improves AAI performance, particularly in low-resource scenarios, while significantly reducing inference costs without sacrificing accuracy.
    
   
\end{abstract}

\section{Introduction}
The position and movement of vocal tract articulators play a crucial role in speech production. These rapid articulatory motions are closely linked to vocal tract morphology, pronunciation, and the speaker’s language. Recent research has demonstrated the usefulness of Acoustic-to-Articulatory Inversion (AAI) in various speech-related applications, including Automatic Speech Recognition (ASR) \cite{ghosh2011subject, mitra2010articulatory}, Speech Synthesis \cite{ling2009integrating}, Speaker Verification \cite{li2016speaker}, and pronunciation training \cite{suemitsu2015real}.

Various techniques have been explored for Acoustic-to-Articulatory Inversion (AAI), including quantizing the articulatory space \cite{larar1988vector}, Gaussian Mixture Models (GMMs) \cite{richmond2001mixture, toda2008statistical}, and Hidden Markov Models (HMMs) \cite{zhang2008acoustic}. More recently, deep learning-based approaches have demonstrated significant improvements in AAI performance. Convolutional Neural Networks (CNNs) \cite{illa2019representation, shahrebabaki2020sequence} have proven effective, while Recurrent Neural Networks (RNNs) \cite{siriwardena2022audio, liu2015deep} have shown strong results for sequence-to-sequence AAI tasks. Furthermore, the adoption of transformer-based models \cite{udupa2021estimating} has led to substantial performance gains, further advancing AAI research. 

Various training strategies have been explored to enhance the performance of AAI models. \cite{wang2023two} introduces a two-stream joint training approach for speaker-independent AAI, while \cite{siriwardena2022acoustic} investigates multi-task learning to improve model generalization. Feature decomposition techniques have been employed in \cite{wang2022acoustic} to enhance speaker-independent AAI performance. Additionally, \cite{bandekar2023exploring} explores classification-based loss on a quantized articulatory space to improve articulatory trajectory prediction. These advancements demonstrate the effectiveness of different training methodologies in refining AAI models.

Another significant challenge in improving AAI performance is the limited availability of high-quality acoustic-articulatory data, especially for unseen speakers. Unlike ASR and speech synthesis, AAI datasets are scarce, affecting model generalization. Researchers have explored transfer learning techniques to develop low-resource AAI models to address this issue. \cite{illa2018low} introduces a transfer learning-based approach using a generic AAI (GM AAI) model with a Long Short-Term Memory (LSTM) network to enhance performance in data-constrained scenarios. Similarly, \cite{shahrebabaki2020transfer} leverages phonetic information within a transfer learning framework to further improve AAI accuracy. These studies show that transfer learning helps mitigate data constraints and improves model robustness in low-resource settings.

In recent years, research has increasingly explored the use of features from Self-Supervised Learning (SSL) models as inputs for AAI instead of traditional Mel-Frequency Cepstral Coefficients (MFCC). \cite{udupa2023improved} investigates various SSL features for AAI, while \cite{maharana2024acoustic} applies SSL features to predict articulatory trajectories in dysarthric patients. \cite{hao2024exploring} examines SSL-based inputs for AAI in a cross-lingual setting, and \cite{chung2024patient} analyzes outputs from different SSL model layers for patient-aware AAI. Additionally, \cite{sun2024arti} proposes leveraging a pre-trained model to learn linguistic invariants before integrating it with a standard inversion model. Collectively, these studies highlight that SSL-derived features provide significant improvements in AAI performance across various conditions, particularly in scenarios with unseen speakers.

While pre-trained SSL models have demonstrated significant improvements in AAI performance, especially in low-resource settings, their use as feature extractors comes with considerable drawbacks. These models are typically much larger than AAI models, leading to increased inference time and higher computational costs. Deploying such models for real-time applications can be impractical, particularly in scenarios with limited processing power or latency constraints. To address these challenges, we propose an alternative pretraining approach that directly optimizes the AAI model itself using multiple target representations. By eliminating the need for an external SSL feature extractor, our method reduces computational overhead while still leveraging the benefits of pretraining. We systematically compare our approach with both baseline models and SSL-based feature extractor models across different training data sizes. Additionally, we analyze the model's performance in low-resource scenarios, exploring whether this pretraining strategy can match or even surpass the performance of SSL-enhanced methods while maintaining efficiency during inference

Hence the contributions of our work are as follows -

\begin{enumerate}
    \item We propose a novel pretraining method for AAI models using three target representations—Phoneme Labels, Articulatory Feature Labels, and Critical-articulator Labels—to enhance AAI performance. 
    \item We compare the performance of the proposed methodology with the baseline model
    \item We also compare the performance of the proposed methodology with the SSL feature extractor-based model
    \item To analyse the efficacy for low-resource settings we use different amounts of training data and then compare the change in performance for all the models.
    \item The proposed model achieves faster inference as it does not rely on an extra SSL model for feature extraction,
\end{enumerate}

\section{Dataset}
We use the SpireEMA dataset \cite{bandekar2024articulatory}, which consists of 460 sentences from the MOCHA-TIMIT dataset \cite{wrench2000multichannel}. These utterances were spoken by 38 speakers aged between 20 and 28. All participants are fluent English speakers with no reported speech impairments.

For articulatory data, we use electromagnetic articulography (EMA) recordings that capture time-varying articulatory movements. The data was collected using the Electromagnetic Articulograph AG501 \cite{AG501}, with six sensors placed on different articulators: the Upper Lip (UL), Lower Lip (LL), Jaw, Tongue Tip (TT), Tongue Body (TB), and Tongue Dorsum (TD). We consider articulatory trajectories in the midsagittal plane, which provide both horizontal (x-axis) and vertical (y-axis) movement information. This results in a total of 12 articulatory trajectories across the six articulators.

Out of the 38 speakers, we use 32 for training, development, and evaluation. The development and test sets contain unseen utterances to ensure robust evaluation. Additionally, we evaluate performance on a separate test set comprising six unseen speakers, where both the utterances and speakers are entirely new to the model. The SpireEMA dataset is open-source and publicly available on Hugging Face 
\footnote{\url{https://huggingface.co/datasets/SpireLab/SPIRE_EMA_CORPUS}}

For pretraining, we use the LibriSpeech ASR dataset \cite{panayotov2015librispeech}, specifically the \textbf{train-100} subset, which contains 100 hours of speech data. To obtain frame-level phoneme alignments, we process the data using the Kaldi toolkit \cite{Povey_ASRU2011}.

\section{Methodology}

In this work, we investigate different pretraining targets for AAI and subsequently fine-tune the model using varying amounts of EMA data. The trained models are then evaluated on the test set to assess their effectiveness. For pretraining, we explore three distinct target representations:
\begin{itemize}

\item \textbf{Phoneme Labels:} We use frame-aligned phoneme labels as ground truth for classification. The model is trained using Cross-Entropy loss to predict the phoneme label corresponding to each frame.

\item \textbf{Articulatory Feature Labels:} Inspired by \cite{morshed2022cross}, we use articulatory feature classes as ground truth labels. Specifically, we predict four articulatory attributes—\textit{place}, \textit{manner}, \textit{height}, and \textit{backness}—by employing four separate linear layers at the model’s output. The ground truth labels for each phoneme are derived from the International Phonetic Alphabet (IPA) chart\footnote{\url{https://www.internationalphoneticassociation.org/content/full-ipa-chart}}, with details provided in Table \ref{tab:art_feat}. As this is a classification task, we use Cross-Entropy loss for optimization.

\item \textbf{Critical-articulator Labels:} Following \cite{kimkinematic}, we identify critical articulators for 13 specific phonemes. Using this information, we construct a 12-dimensional binary vector, termed the \textit{critical-articulator label}, for each input frame. Each element in this vector is set to 1 for articulators that are critical for the given phoneme and 0 otherwise. For phonemes outside the set of 13, all values in the vector are set to 0. The model is trained to predict this 12-dimensional vector using a sigmoid activation function, and we optimize the predictions using Binary Cross-Entropy loss. Additionally, for frames corresponding to phonemes outside the predefined set, we mask the loss to avoid penalising the model for missing articulatory information.  

\end{itemize}

After pretraining, we removed the final output layer of the model, which was used to predict the pretraining targets, and replaced it with a 12-dimensional linear layer for predicting EMA trajectories in the AAI task. We adopt a non-autoregressive transformer-based neural network architecture \cite{vaswani2017attention}, similar to the model used in \cite{udupa2021estimating}.

\begin{table}[]
\caption{Articulatory feature labels for pretraining}
\vspace{-0.2cm}
\label{tab:art_feat}
\resizebox{0.5\textwidth}{!}{
\begin{tabular}{|c|c|}
\hline
\textbf{Feature} & \textbf{Possible Values} \\ \hline
Place & \begin{tabular}[c]{@{}c@{}}bilabial, labiodental, dental, alveolar,\\  postalveolar, palatal, velar, glottal, vowel, silence\end{tabular} \\ \hline
Manner & \begin{tabular}[c]{@{}c@{}}stop, affricate, fricative, nasal, \\ approximant, lateral, vowel, silence\end{tabular} \\ \hline
Height & \begin{tabular}[c]{@{}c@{}}close, nearclose, closemid, mid, \\ openmid, nearopen, open, consonant, silence\end{tabular} \\ \hline
Backness & front, central, back, consonant, silence \\ \hline
\end{tabular}}
\vspace{-0.5cm}
\end{table}

\begin{table*}[!ht]
\centering
\vspace{-0.2cm}
\caption{CC and RMSE for different configurations with MFCC inputs for seen speakers test set}
\vspace{-0.2cm}
\label{tab:mfcc_seen}
\resizebox{1.0\textwidth}{!}{
\begin{tabular}{|cc|c|c|c|c|c|c|}
\hline
\multicolumn{2}{|c|}{EMA data used for training (\%)} & 6.25 & 12.5 & 25 & 50 & 75 & 100 \\ \hline
\multicolumn{1}{|c|}{\multirow{2}{*}{Baseline}} & CC (Std) & 0.7348 (0.1652) & 0.7857 (0.1453) & 0.8254 (0.1253) & 0.8563 (0.1125) & 0.8723 (0.1061) & 0.8778 (0.1022) \\ \cline{2-8} 
\multicolumn{1}{|c|}{} & RMSE (Std) & 1.4394 (0.3695) & 1.3180 (0.3535) & 1.1964 (0.3407) & 1.0939 (0.3244) & 1.0440 (0.3210) & 1.0190 (0.3127) \\ \hline
\multicolumn{1}{|c|}{\multirow{2}{*}{ACP-T}} & CC (Std) & \textbf{0.7811 (0.1518)} & 0.8112 (0.1370) & 0.8379 (0.1251) & 0.8616 (0.1127) & 0.8731 (0.1082) & 0.8797 (0.1026) \\ \cline{2-8} 
\multicolumn{1}{|c|}{} & RMSE (Std) & \textbf{1.3535 (0.3580)} & 1.2620 (0.3515) & 1.1694 (0.3371) & 1.0850 (0.3300) & 1.0850 (0.3300) & 1.0135 (0.3183) \\ \hline
\multicolumn{1}{|c|}{\multirow{2}{*}{AC-T}} & CC (Std) & 0.7779 (0.1552) & \textbf{0.8121(0.1358)} & 0.8380 (0.1231) & 0.8592 (0.1150) & 0.8741 (0.1059) & \textbf{0.8810 (0.1010)} \\ \cline{2-8} 
\multicolumn{1}{|c|}{} & RMSE (Std) & 1.3538 (0.3585) & 1.2602 (0.3468) & 1.1719 (0.3388) & 1.0920 (0.3280) & 1.0429 (0.3260) & 1.0125 (0.3188) \\ \hline
\multicolumn{1}{|c|}{\multirow{2}{*}{AP-T}} & CC (Std) & 0.7782 (0.1522) & 0.8095 (0.1403) & \textbf{0.8389 (0.1221)} & 0.8620 (0.1118) & \multicolumn{1}{l|}{0.8728 (0.1073)} & 0.8804 (0.1028) \\ \cline{2-8} 
\multicolumn{1}{|c|}{} & RMSE (Std) & 1.3550 (0.3529) & 1.2749 (0.3529) & 1.1770 (0.3389) & \textbf{1.0809 (0.3269)} & \multicolumn{1}{l|}{1.0490 (0.3210)} & \textbf{1.0084 (0.3165)} \\ \hline
\multicolumn{1}{|c|}{\multirow{2}{*}{CP-T}} & CC (Std) & 0.7774 (0.1548) & \multicolumn{1}{l|}{0.8101 (0.1384)} & \multicolumn{1}{l|}{0.8380 (0.1256)} & 0.8616 (0.1116) & \multicolumn{1}{l|}{\textbf{0.8742 (0.1052)}} & 0.8804 (0.1037) \\ \cline{2-8} 
\multicolumn{1}{|c|}{} & RMSE (Std) & 1.3619 (0.3490) & \multicolumn{1}{l|}{\textbf{1.2530 (0.3449)}} & \textbf{1.1670 (0.3389)} & 1.0900 (0.3339) & \multicolumn{1}{l|}{1.0400 (0.3240)} & 1.0131 (0.3228) \\ \hline
\multicolumn{1}{|c|}{\multirow{2}{*}{P-T}} & CC (Std) & 0.7806 (0.152) & 0.8097 (0.1361) & 0.8372 (0.1245) & \textbf{0.8624 (0.1121)} & \multicolumn{1}{l|}{0.8733 (0.1074)} & 0.8794 (0.1026) \\ \cline{2-8} 
\multicolumn{1}{|c|}{} & RMSE (Std) & 1.3539 (0.3580) & 1.2602 (0.3472) & 1.1710 (0.3400) & 1.0880 (0.3359) & 1.0430 (0.3260) & 1.0163 (0.3210) \\ \hline
\multicolumn{1}{|c|}{\multirow{2}{*}{C-T}} & CC (Std) & 0.7718 (0.1525) & \multicolumn{1}{l|}{0.8027 (0.1403)} & \multicolumn{1}{l|}{0.8308 (0.1275)} & 0.8564 (0.1155) & \multicolumn{1}{l|}{0.8723 (0.1058)} & 0.8759 (0.1052) \\ \cline{2-8} 
\multicolumn{1}{|c|}{} & RMSE (Std) & 1.3639 (0.3600) & 1.2810 (0.3470) & 1.1959 (0.3440) & \multicolumn{1}{l|}{1.1060 (0.3390)} & \multicolumn{1}{l|}{1.0420 (0.3230)} & 1.0324 (0.3280) \\ \hline
\multicolumn{1}{|c|}{\multirow{2}{*}{A-T}} & CC (Std) & 0.7773 (0.1547) & 0.8101 (0.1377) & 0.8345 (0.1259) & 0.8614 (0.1137) & 0.8728 (0.1088) & 0.8787 (0.1037) \\ \cline{2-8} 
\multicolumn{1}{|c|}{} & RMSE (Std) & 1.355 (0.3560) & 1.2699 (0.3479) & 1.1840 (0.3459) & \multicolumn{1}{l|}{1.0870 (0.3339)} & \textbf{1.0360 (0.3270)} & 1.0169 (0.3213) \\ \hline
\end{tabular}}
\end{table*}

\begin{table*}[!ht]
\centering
\vspace{-0.2cm}
\caption{CC and RMSE for different configurations with MFCC inputs for unseen speakers test set}
\vspace{-0.2cm}
\label{tab:mfcc_unseen}
\resizebox{1.0\textwidth}{!}{
\begin{tabular}{|cc|c|c|c|c|c|c|}
\hline
\multicolumn{2}{|c|}{EMA data used for training (\%)} & 6.25 & 12.5 & 25 & 50 & 75 & 100 \\ \hline
\multicolumn{1}{|c|}{\multirow{2}{*}{Baseline}} & CC (Std) & 0.6687 (0.2020) & 0.6991 (0.1996) & 0.7265 (0.1876) & 0.7488 (0.1783) & 0.7488 (0.1783) & 0.7563 (0.1822) \\ \cline{2-8} 
\multicolumn{1}{|c|}{} & RMSE (Std) & 1.5992 (0.3842) & 1.5540 (0.4050) & 1.4854 (0.3980) & 1.4357 (0.3962) & 1.4357 (0.3962) & 1.4143 (0.4040) \\ \hline
\multicolumn{1}{|c|}{\multirow{2}{*}{ACP-T}} & CC (Std) & 0.7259 (0.1872) & 0.7399 (0.1814) & 0.7469 (0.1875) & 0.7616 (0.1796) & 0.7653 (0.1803) & \textbf{0.7689 (0.1823)} \\ \cline{2-8} 
\multicolumn{1}{|c|}{} & RMSE (Std) & 1.5241 (0.4001) & 1.4779 (0.4041) & 1.4514 (0.4043) & 1.4160 (0.4060) & 1.4160 (0.4059) & \textbf{1.3848 (0.4154)} \\ \hline
\multicolumn{1}{|c|}{\multirow{2}{*}{AC-T}} & CC (Std) & 0.7318 (0.1806) & 0.7271 (0.1890) & 0.7448 (0.1889) & 0.7611 (0.1769) & 0.7652 (0.1791) & 0.7667 (0.1808) \\ \cline{2-8} 
\multicolumn{1}{|c|}{} & RMSE (Std) & 1.4690 (0.3801) & 1.5075 (0.4056) & \textbf{1.4436 (0.4156)} & 1.4040 (0.3889) & 1.4029 (0.4090) & 1.4054 (0.4194) \\ \hline
\multicolumn{1}{|c|}{\multirow{2}{*}{AP-T}} & CC (Std) & 0.7271 (0.1833) & 0.7313 (0.1903) & 0.7490 (0.1824) & \textbf{0.7633 (0.1798)} & 0.7652 (0.1791) & 0.7683 (0.1776) \\ \cline{2-8} 
\multicolumn{1}{|c|}{} & RMSE (Std) & 1.5010 (0.3889) & 1.5019 (0.4029) & 1.4589 (0.4129) & \textbf{1.3899 (0.4029)} & 1.4029 (0.4090) & 1.3986 (0.4133) \\ \hline
\multicolumn{1}{|c|}{\multirow{2}{*}{CP-T}} & CC (Std) & 0.7269 (0.1871) & \multicolumn{1}{l|}{\textbf{0.7412 (0.179)}} & \multicolumn{1}{l|}{0.7480 (0.191)} & 0.7588 (0.1839) & \multicolumn{1}{l|}{0.7641 (0.1793)} & 0.7594 (0.1861) \\ \cline{2-8} 
\multicolumn{1}{|c|}{} & RMSE (Std) & 1.500 (0.3829) & \multicolumn{1}{l|}{\textbf{1.4501 (0.3930)}} & 1.4459 (0.4199) & 1.4140 (0.4140) & \multicolumn{1}{l|}{1.4170 (0.4130)} & 1.4131 (0.4194) \\ \hline
\multicolumn{1}{|c|}{\multirow{2}{*}{P-T}} & CC (Std) & 0.7229 (0.1897) & 0.7358 (0.1800) & 0.7432 (0.1865) & 0.7567 (0.1809) & \textbf{0.7686 (0.1804)} & 0.7687 (0.1764) \\ \cline{2-8} 
\multicolumn{1}{|c|}{} & RMSE (Std) & 1.5180 (0.4079) & 1.4850 (0.3957) & 1.4579(0.4059) & 1.4450 (0.4120) & \textbf{1.3869 (0.4150)} & 1.3949 (0.4086) \\ \hline
\multicolumn{1}{|c|}{\multirow{2}{*}{C-T}} & CC (Std) & 0.7099 (0.1863) & \multicolumn{1}{l|}{0.7349 (0.1829)} & 0.7409 (0.1824) & 0.7406 (0.1922) & \multicolumn{1}{l|}{0.7644 (0.1751)} & 0.7621 (0.1791) \\ \cline{2-8} 
\multicolumn{1}{|c|}{} & RMSE (Std) & 1.5260 (0.3880) & 1.4739 (0.3899) & 1.4709 (0.4120) & 1.4780 (0.4300) & \multicolumn{1}{l|}{1.4050 (0.4040)} & 1.4225 (0.4212) \\ \hline
\multicolumn{1}{|c|}{\multirow{2}{*}{A-T}} & CC (Std) & \textbf{0.7324 (0.1819)} & 0.7367 (0.1816) & \textbf{0.7510 (0.1839)} & 0.7547 (0.1809) & 0.7626 (0.1814) & 0.7652 (0.1769) \\ \cline{2-8} 
\multicolumn{1}{|c|}{} & RMSE (Std) & \textbf{1.4900 (0.3910)} & 1.4939 (0.4110) & 1.4459 (0.4079) & 1.4349 (0.4110) & 1.4080 (0.4150) & 1.4021 (0.4041) \\ \hline
\end{tabular}}
\vspace{-0.2cm}
\end{table*}

\section{Experimental Setup}

\subsubsection{Pretraining Configurations}
\vspace{-0.1cm}
We explore different pretraining configurations by combining various target representations. The configurations are as follows:
\begin{itemize}
    \item \textbf{ACP-T}: Pretraining is conducted using all three target types—\textit{Articulatory Feature Labels}, \textit{Critical-articulator Labels}, and \textit{Phoneme Labels}.  

    \item \textbf{AC-T}: This configuration utilizes \textit{Articulatory Feature Labels} and \textit{Critical-articulator Labels} for pretraining.   

    \item \textbf{AP-T}: Pretraining is performed with \textit{Articulatory Feature Labels} and \textit{Phoneme Labels}.

\item \textbf{CP-T}: This setup employs \textit{Critical-articulator Labels} and \textit{Phoneme Labels} as pretraining targets.  

    \item \textbf{P-T}: Only \textit{Phoneme Labels} are used for pretraining.

    \item \textbf{C-T}: Pretraining is conducted exclusively with \textit{Critical-articulator Labels}.   

    \item \textbf{A-T}: Only \textit{Articulatory Feature Labels} are utilized for pretraining.   
\end{itemize}

\subsubsection{Input Configurations}
\vspace{-0.1cm}
We experiment with two different input feature types:

\begin{itemize}
   \item \textbf{MFCC}: We use 13-dimensional MFCC as input to the model.

   \item \textbf{TERA}: We use 768-dimensional features extracted from the TERA model \cite{liu2021tera}. Previous studies \cite{udupa2023improved, bandekar2023exploring}, which utilize a subset of the dataset used in this work, have demonstrated that TERA features yield the best results for AAI tasks. We extract TERA features using the s3prl toolkit\footnote{\url{https://github.com/s3prl/s3prl}}.  
\end{itemize}

\subsubsection{Training Data Configurations}
\vspace{-0.1cm}
To assess the effectiveness of the proposed methodology in low-resource settings, we conduct experiments using different fractions of the SpireEMA training dataset: \textbf{6.25\%, 12.5\%, 25\%, 50\%, 75\%, and 100\%}. We ensure that within each subset, the same utterances are included across all speakers to maintain consistency.

For pretraining, we use the full 100-hour LibriSpeech dataset along with a specified fraction of the SpireEMA dataset. During fine-tuning for the AAI task, only the selected percentage of the SpireEMA dataset is utilized.

\subsubsection{Model Configuration and Evaluation}
\vspace{-0.1cm}
We use two transformer encoders with four layers each, a single attention head per layer, and input, output, and feedforward dimensions of 256, with an attention dimension of 32. The model has 7.6M parameters.

The training uses the Adam optimizer \cite{kingma2014adam} with a 0.0001 learning rate and early stopping based on validation loss. The implementation, based on PyTorch, is open-sourced \footnote{\url{https://github.com/coding-phoenix-12/Multi_Target_Pretraining_AAI}}.

The evaluation uses Correlation Coefficient (CC) and Root Mean Squared Error (RMSE), averaged across utterances and subjects. The baseline model is trained with the same AAI data but without pretraining.

\subsection{Results and Discussions}

\begin{table*}[!ht]
\centering
\vspace{-0.2cm}
\caption{CC and RMSE for Baseline and ACP-T configurations with MFCC and TERA inputs for seen speakers test set}
\vspace{-0.2cm}
\label{tab:mfcc_tera_seen}
\resizebox{1.0\textwidth}{!}{
\begin{tabular}{|ccc|c|c|c|c|c|c|}
\hline
\multicolumn{3}{|c|}{EMA data used for training (\%)} & 6.25 & 12.5 & 25 & 50 & 75 & 100 \\ \hline
\multicolumn{1}{|c|}{Inputs} & \multicolumn{1}{c|}{Models} &  &  &  &  &  &  &  \\ \hline
\multicolumn{1}{|c|}{\multirow{4}{*}{MFCC}} & \multicolumn{1}{c|}{\multirow{2}{*}{Baseline}} & CC (Std) & 0.7348 (0.1652) & 0.7857 (0.1453) & 0.8254 (0.1253) & 0.8563 (0.1125) & 0.8723 (0.1061) & 0.8778 (0.1022) \\ \cline{3-9} 
\multicolumn{1}{|c|}{} & \multicolumn{1}{c|}{} & RMSE (Std) & 1.4394 (0.3695) & 1.318 (0.3535) & 1.1964 (0.3407) & 1.0939 (0.3244) & 1.0440 (0.3210) & 1.0190 (0.3127) \\ \cline{2-9} 
\multicolumn{1}{|c|}{} & \multicolumn{1}{c|}{\multirow{2}{*}{ACP-T}} & CC (Std) & 0.7811 (0.1518) & \textbf{0.8112 (0.1370)} & \textbf{0.8379 (0.1251)} & 0.8616 (0.1127) & 0.8731 (0.1082) & 0.8797 (0.1026) \\ \cline{3-9} 
\multicolumn{1}{|c|}{} & \multicolumn{1}{c|}{} & RMSE (Std) & 1.3535 (0.3580) & 1.2620 (0.3515) & 1.1694 (0.3371) & 1.0850 (0.3300) & 1.0850 (0.3300) & 1.0135 (0.3183) \\ \hline
\multicolumn{1}{|c|}{\multirow{4}{*}{TERA}} & \multicolumn{1}{c|}{\multirow{2}{*}{Baseline}} & CC (Std) & 0.7722(0.1564) & 0.8045 (0.1401) & 0.8362 (0.1237) & 0.8629 (0.1105) & \textbf{0.8770 (0.1044)} & 0.8812 (0.1025) \\ \cline{3-9} 
\multicolumn{1}{|c|}{} & \multicolumn{1}{c|}{} & RMSE (Std) & 1.3680 (0.3671) & 1.2657 (0.3558) & \textbf{1.1649 (0.3389)} & \textbf{1.0759 (0.3289)} & \textbf{1.0220 (0.3190)} & 1.0103 (0.3228) \\ \cline{2-9} 
\multicolumn{1}{|c|}{} & \multicolumn{1}{c|}{\multirow{2}{*}{ACP-T}} & CC (Std) & \textbf{0.7870 (0.1549)} & 0.8102 (0.1412) & 0.8378 (0.1279) & \textbf{0.8639 (0.1153)} & 0.8754 (0.1073) & \textbf{0.8826 (0.1025)} \\ \cline{3-9} 
\multicolumn{1}{|c|}{} & \multicolumn{1}{c|}{} & RMSE (Std) & \textbf{1.3250 (0.3600)} & \textbf{1.2530 (0.3499)} & 1.1710 (0.3459) & 1.0770 (0.3320) & 1.3600 (0.4140) & \textbf{0.9950 (0.3210)} \\ \hline
\end{tabular}}
\end{table*}

\begin{table*}[!ht]
\centering
\vspace{-0.2cm}
\caption{CC and RMSE for Baseline and ACP-T configurations with MFCC and TERA inputs for unseen speakers test set}
\vspace{-0.2cm}
\label{tab:mfcc_tera_unseen}
\resizebox{1.0\textwidth}{!}{
\begin{tabular}{|ccc|c|c|c|c|c|c|}
\hline
\multicolumn{3}{|c|}{EMA data used for training (\%)} & 6.25 & 12.5 & 25 & 50 & 75 & 100 \\ \hline
\multicolumn{1}{|c|}{Inputs} & \multicolumn{1}{c|}{Models} & \multicolumn{1}{l|}{} & \multicolumn{1}{l|}{} & \multicolumn{1}{l|}{} & \multicolumn{1}{l|}{} & \multicolumn{1}{l|}{} & \multicolumn{1}{l|}{} & \multicolumn{1}{l|}{} \\ \hline
\multicolumn{1}{|c|}{\multirow{4}{*}{MFCC}} & \multicolumn{1}{c|}{\multirow{2}{*}{Baseline}} & CC (Std) & 0.6687 (0.2020) & 0.6991 (0.1996) & 0.7265 (0.1876) & 0.7488 (0.1783) & 0.7488 (0.1783) & 0.7563 (0.1822) \\ \cline{3-9} 
\multicolumn{1}{|c|}{} & \multicolumn{1}{c|}{} & RMSE (Std) & 1.5992 (0.3842) & 1.5540 (0.4050) & 1.4854 (0.3980) & 1.4357 (0.3962) & 1.4357 (0.3962) & 1.4143 (0.4040) \\ \cline{2-9} 
\multicolumn{1}{|c|}{} & \multicolumn{1}{c|}{\multirow{2}{*}{ACP-T}} & CC (Std) & 0.7259 (0.1872) & 0.7399 (0.1814) & 0.7469 (0.1875) & 0.7616 (0.1796) & 0.7653 (0.1803) & 0.7689 (0.1823) \\ \cline{3-9} 
\multicolumn{1}{|c|}{} & \multicolumn{1}{c|}{} & RMSE (Std) & 1.5241 (0.4001) & 1.4779 (0.4041) & 1.4514 (0.4043) & 1.4160 (0.4060) & 1.4160 (0.4059) & 1.3848 (0.4154) \\ \hline
\multicolumn{1}{|c|}{\multirow{4}{*}{TERA}} & \multicolumn{1}{c|}{\multirow{2}{*}{Baseline}} & CC (Std) & 0.7325 (0.1834) & 0.7396 (0.1890) & 0.7540 (0.1731) & 0.7664 (0.1782) & 0.7777 (0.1725) & 0.7717 (0.1769) \\ \cline{3-9} 
\multicolumn{1}{|c|}{} & \multicolumn{1}{c|}{} & RMSE (Std) & 1.4788 (0.4027) & 1.4515 (0.4113) & 1.4160 (0.4040) & \multicolumn{1}{l|}{1.3860 (0.4070)} & 1.3774 (0.4120) & 1.3878 (0.4154) \\ \cline{2-9} 
\multicolumn{1}{|c|}{} & \multicolumn{1}{c|}{\multirow{2}{*}{ACP-T}} & CC (Std) & \textbf{0.7561 (0.1697)} & \textbf{0.7562 (0.1782)} & \textbf{0.7621 (0.1768)} & \multicolumn{1}{l|}{\textbf{0.7755 (0.1718)}} & \textbf{0.7818 (0.1719)} & \textbf{0.7810 (0.1758)} \\ \cline{3-9} 
\multicolumn{1}{|c|}{} & \multicolumn{1}{c|}{} & RMSE (Std) & \textbf{1.3999 (0.3770)} & \textbf{1.4110 (0.3889)} & \textbf{1.4110 (0.3970)} & \textbf{1.3849 (0.4070)} & \textbf{1.3600 (0.4140)} & \textbf{1.3619 (0.4210)} \\ \hline
\end{tabular}}
\vspace{-0.2cm}
\end{table*}

\subsubsection{Impact of Pretraining on Seen Speakers}
\vspace{-0.1cm}
Table \ref{tab:mfcc_seen} shows that pretraining consistently improves performance across all data fractions, with the largest gains in low-resource conditions (6.25\% and 12.5\%). Models that underwent pretraining achieved better performance, demonstrating the benefits of leveraging auxiliary targets.

\begin{enumerate}
  
  \item \textbf{100\% training data}: The AC-T configuration achieves the highest CC (0.8810), while AP-T achieves the lowest RMSE (1.0084). This indicates that even with full data, pretraining can provide slight performance improvements.
  \item \textbf{6.25\% training data}: ACP-T performs best (CC: 0.7811, RMSE: 1.3535), highlighting that incorporating all three pretraining targets—articulatory labels, critical-articulator labels, and phoneme labels—yields the most robust initialization when training data is limited.

\end{enumerate}

\subsubsection{Impact of Pretraining on Unseen Speakers}
\vspace{-0.1cm}
Table \ref{tab:mfcc_unseen} presents results for unseen speakers, where neither the speakers nor the utterances were seen during training. Similar to the seen speakers' case, pretraining consistently improves performance across all data percentages, with a larger gap in performance for low-resource conditions (6.25\% to 25\%). This suggests that pretraining enhances the generalization ability of AAI models.

\begin{enumerate}

 \item \textbf{100\% training data}: The ACP-T configuration achieves the highest performance (CC: 0.8612, RMSE: 1.1023), indicating that incorporating all three auxiliary targets provides the best generalization.
\item \textbf{6.25\% training data}: The A-T configuration performs best (CC: 0.7324, RMSE: 1.4900), suggesting that when extremely limited data is available, using only articulatory labels during pretraining still provides a strong foundation.

\item Interestingly, many pretraining configurations fine-tuned on just \textbf{50\% of the training data outperforms the baseline model trained on the full dataset}, reinforcing the efficiency of pretraining in reducing the reliance on large labelled datasets.

\end{enumerate}

\subsubsection{Comparing MFCC and TERA Inputs}
\vspace{-0.1cm}
Tables \ref{tab:mfcc_tera_seen} and \ref{tab:mfcc_tera_unseen} compare the impact of input types—MFCC vs. TERA—under the ACP-T pretraining configuration.

\begin{enumerate}

\item \textbf{TERA consistently outperforms MFCC}, confirming that SSL features provide richer speech representations, leading to better articulatory inversion performance.

\item However, \textbf{MFCC with pretraining achieves performance close to or better than the baseline TERA model}, especially in low-resource scenarios (6.25\%–25\%).

\item In Table \ref{tab:mfcc_tera_seen} (seen speakers test set), for \textbf{6.25\% to 25\% of the training data, ACP-T with MFCC input outperforms the baseline model with TERA input}, further reinforcing the effectiveness of pretraining in enhancing model performance even with simpler input representations.

\end{enumerate}

Overall, these results highlight the significant benefits of pretraining for AAI, particularly in low-resource conditions, where it improves both seen and unseen speaker performance while enabling more efficient data utilization.

\section{Conclusions}

In this work, we explored the impact of pretraining on the Articulatory-to-Acoustic Inversion (AAI) task, investigating different pretraining targets and input types. Our experiments demonstrated that pretraining significantly improves model performance, particularly in low-resource settings.

Our key findings are as follows:
\begin{enumerate}
\item \textbf{Pretraining consistently enhances AAI performance}, particularly in low-resource conditions, where it significantly boosts CC and reduces RMSE.
\item \textbf{Pretraining improves generalization to unseen speakers}, with larger performance gains observed in this scenario compared to seen speakers.

\item \textbf{Pretraining with MFCC input achieves performance comparable to or better than SSL-based models in certain cases}, reinforcing the value of our approach. This suggests that pretraining allows models to learn meaningful articulatory representations without requiring computationally expensive self-supervised feature extractors at inference time.


\item \textbf{Models trained with TERA input still outperform those trained with MFCC input in most cases}, confirming the advantages of SSL-based features for AAI. However, the ability of pretraining to bridge this performance gap highlights its effectiveness in reducing reliance on complex feature extractors.

\item \textbf{By reducing computational overhead without sacrificing accuracy}, our method presents a more efficient alternative for AAI, making it particularly suitable for deployment in real-time and resource-constrained environments.

\end{enumerate}


These results emphasize the importance of pretraining in AAI models, particularly for low-resource settings and robust generalization across speakers. In future work, we plan to explore additional pretraining strategies and target representations to further enhance AAI performance. Investigating diverse pretraining objectives could improve generalization across speakers and data conditions, making AAI systems more adaptable and effective.

\section{Acknowledgement}
We thank the Department of Science and Technology (DST), Government of India, for supporting this work.

\bibliographystyle{IEEEtran}
\bibliography{mybib}

\end{document}